\documentclass{elsart}

\def\simgt{\lower 2pt \hbox{$\, \buildrel {\scriptstyle >}\over {\scriptstyle \sim}\,$}}
\def\simlt{\lower 2pt \hbox{$\, \buildrel {\scriptstyle <}\over {\scriptstyle \sim}\,$}}

\def\chandra{{\it Chandra\/}}
\def\conx{{\it Constellation-X\/}}
\def\duo{{\it DUO\/}}

\def\exist{{\it EXIST\/}}
\def\genx{{\it Generation-X\/}}
\def\hst{{\it {\it HST}\/}}
\def\jdem{{\it {\it JDEM}\/}}
\def\nustar{{\it {\it NuSTAR}\/}}
\def\rosat{{\it ROSAT\/}}

\def\xeus{{\it XEUS\/}}
\def\xmm{{\it XMM-Newton\/}}

\begin{document}


\begin{frontmatter}

\title{X-ray Surveys and Wide-Field Optical/Near-Infrared Imaging with JDEM}

\author{W.N. Brandt}
\address{Department of Astronomy \& Astrophysics, The Pennsylvania 
State University, 525 Davey Lab, University Park, PA 16802}

\begin{abstract}
I briefly describe a few important scientific issues that could 
be addressed effectively via the combination of data from \jdem\ and 
\hbox{X-ray} missions. The topics covered are largely focused on active 
galactic nuclei (AGN) and include
(1) the selection of AGN via \hbox{X-ray} emission and optical variability, 
(2) nuclear outbursts in galaxies due to transient fueling
of their supermassive black holes, 
(3) moderate-luminosity AGN at high redshift ($z>4$) found via 
application of ``dropout'' techniques to X-ray sources, and
(4) the host-galaxy morphologies of \hbox{X-ray} selected AGN. 
I also describe the substantial challenges to obtaining wide-field 
\hbox{X-ray} data with sufficient sensitivity to complement \jdem\ properly. 
\end{abstract}

\end{frontmatter}


\section{Introduction}

Extragalactic \hbox{X-ray} surveys have dramatically advanced over the past 
five years, largely due to the flood of data from the {\it Chandra X-ray 
Observatory\/} (hereafter \chandra) and the {\it X-ray Multi-Mirror 
Mission-Newton\/} (hereafter \xmm). The superb \hbox{X-ray} mirrors and 
charge-coupled device (CCD) detectors on these observatories provide
large source samples detected to \hbox{$\approx 0.5$--10~keV} flux 
levels that are up to \hbox{50--250} times lower than those of previous \hbox{X-ray} 
missions. They furthermore provide high-quality \hbox{X-ray} source positions 
with accuracies of \hbox{$\approx$~0.3--3$^{\prime\prime}$}, allowing
reliable matching to multiwavelength counterparts. About 40 extragalactic
surveys are presently underway utilizing data from \chandra\ and \xmm.

While a broad diversity of sources are found in \chandra\ and \xmm\ 
surveys, it is clear that the majority of the detected sources are 
active galactic nuclei (AGN). In fact, deep \chandra\ and \xmm\ surveys 
have found the highest AGN sky densities to date, reaching 
$\approx 7,200$~deg$^{-2}$ (e.g., Bauer et~al. 2004). \hbox{X-ray} surveys
appear to have excellent, although not perfect, completeness when 
compared to other multiwavelength methods for finding AGN. 

Many of the AGN found in \chandra\ and \xmm\ surveys are faint at optical 
and other wavelengths, often having $I$ magnitudes of 23--27 (see Fig.~1). 
They also often have modest optical luminosities so that, without 
superb imaging from space, there is substantial blending of the light 
from the AGN and its host galaxy. Thus, sensitive and high-resolution 
imaging at multiple wavelengths and over relatively large fields is 
required for optimal follow-up studies.\footnote{The effective field 
of view for \chandra\ (\xmm) is \hbox{$\sim 1/12$~deg$^2$} 
\hbox{($\sim 1/5$~deg$^2$)}, and most of the surveys mentioned above 
use \hbox{1--20} such fields.} Such imaging has typically lagged 
behind the \hbox{X-ray} data due to observational expense, but the 
{\it Joint Dark Energy Mission\/} (\jdem) would dramatically reverse this 
situation, likely leaving \hbox{X-ray} astronomers scrambling to ``catch up.'' 

\begin{figure}
\includegraphics{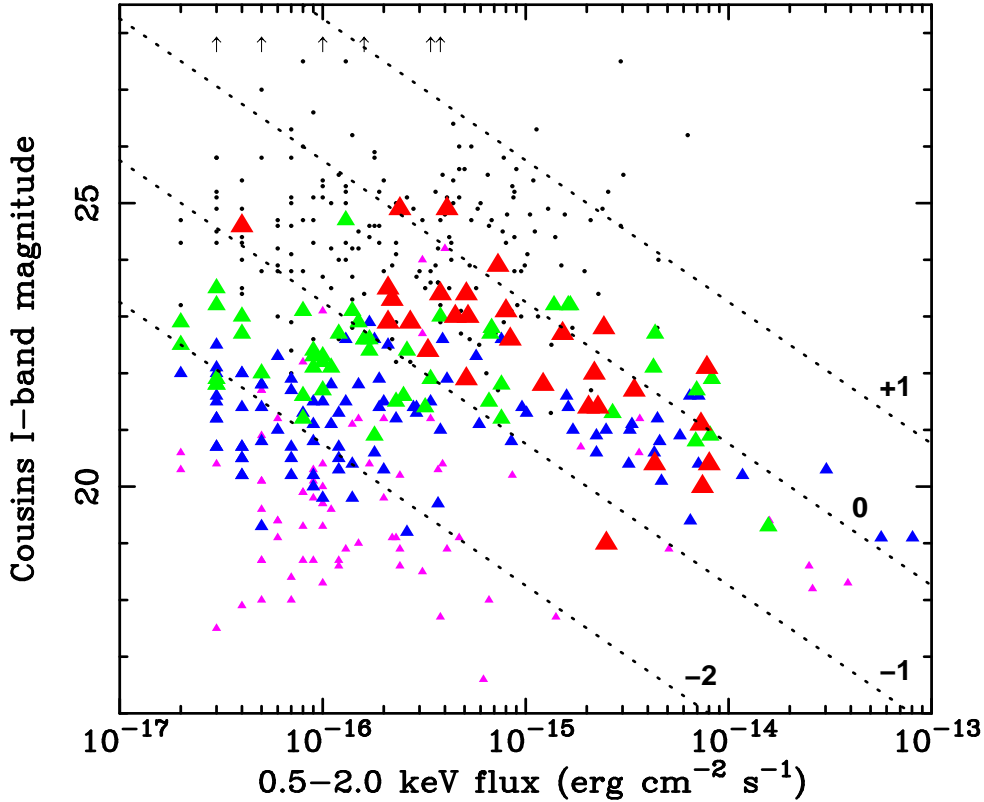}
\vspace{4.3in}
\caption{$I$-band magnitude versus \hbox{0.5--2~keV} flux for extragalactic 
\hbox{X-ray} sources in the 2~Ms Chandra Deep Field-North (CDF-N) observation. 
Sources with redshifts of \hbox{0--0.5}, \hbox{0.5--1}, \hbox{1--2}, 
and \hbox{2--6} are shown as violet, blue, green, and 
red filled triangles, respectively (symbol sizes also increase with redshift). 
Small black dots indicate sources without measured redshifts. The slanted, dotted
lines indicate constant values of $\log (f_{\rm X}/f_{\rm I})$; the respective
$\log (f_{\rm X}/f_{\rm I})$ values are labeled.}
\end{figure} 

Below I will briefly describe a few exciting science projects that
would be enabled by combining data from \jdem\ and \hbox{X-ray} missions such
as \chandra, \conx, \duo, \exist, \genx, \nustar, \xeus, and \xmm. 
I will focus on AGN-driven science as AGN are the numerically
dominant source population in extragalactic \hbox{X-ray} surveys. However, 
wonderful \jdem/\hbox{X-ray} science should also be possible for 
clusters and groups of galaxies, large-scale structures, starburst 
galaxies, normal galaxies, and other classes of objects. 


\section{The Selection of AGN via \hbox{X-ray} Emission and Optical Variability}

Deep optical variability surveys are one of the only methods competitive 
with \hbox{X-ray} surveys at efficiently finding high sky densities of 
AGN. For example, Sarajedini et~al. (2003) report the discovery of 16 
variable galactic nuclei in the Hubble Deep Field-North (HDF-N) based 
on analyses of two \hst\ $V$-band images taken five years apart
(see Fig.~2). The 
derived AGN sky density is in the range
$\approx 2,000$--11,000~deg$^{-2}$. The 
precise sky density depends upon the amount of contamination by 
nuclear supernovae and statistical outliers, but it is clearly higher 
than that from most optical spectroscopic surveys and is plausibly
comparable to that from the deepest \hbox{X-ray} surveys (see \S1). 

\begin{figure}
\includegraphics{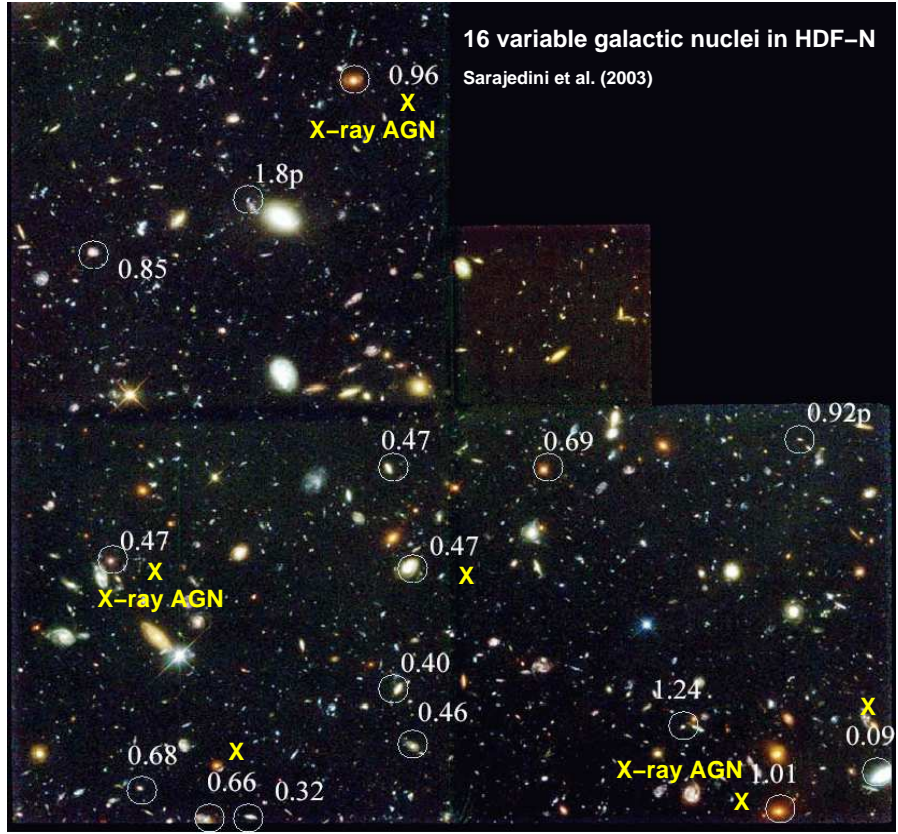}
\vspace{4.5in}
\caption{\hst\ image of the HDF-N showing the 16 galaxies with optically
variable nuclei (circled and labeled with redshifts). The galaxies with
both variable nuclei and X-ray detections are marked with ``X'' symbols;
those that are clear X-ray AGN are labeled as such (there are additional
X-ray AGN in the HDF-N that have not shown variable nuclei). Adapted from
Alexander et al. (2003) and Sarajedini et~al. (2003).}
\end{figure} 

While \hbox{X-ray} selection and optical variability selection have respectable 
overlap in the AGN found (see Fig.~2), each method also finds AGN missed by 
the other. For example, there are at least 2--3 \hbox{X-ray} AGN in the 
HDF-N not identified by Sarajedini et~al. (2003). Conversely, at 
least some of the variable galactic nuclei of Sarajedini et~al. (2003)
may be \hbox{X-ray} weak AGN lying below the detection threshold of even the
2~Ms CDF-N exposure. These AGN might be \hbox{X-ray} weak due to a low 
bolometric power, or they might have emaciated accretion-disk coronae
that cannot produce \hbox{X-rays} effectively. 

As currently planned, \jdem\ should find at least 30,000--90,000
variability selected AGN in its 15~deg$^2$ deep survey. An even
larger number of AGN should be found in its wider surveys. If
appropriately matched \hbox{X-ray} data were obtained (see \S6), joint
\jdem\ and \hbox{X-ray} techniques could be applied to generate the largest
and most complete census of moderate-luminosity, typical AGN out
to high redshift. \hbox{X-ray} sources could be scrutinized especially
carefully for optically variable nuclei in the \jdem\ data, and
\hbox{X-ray} stacking techniques could be used to study the average \hbox{X-ray} 
properties of AGN not detected individually. The large solid angle
coverage would allow the AGN population to be probed consistently
over a wide range of luminosity; the present data have poor coverage
of luminous, rare AGN. 


\section{Nuclear Outbursts in Galaxies}

X-ray surveys, mainly with \rosat, have discovered about seven 
large-amplitude \hbox{X-ray} outbursts from galactic nuclei
(e.g., Donley et~al. 2002; Komossa 2002; Vaughan et~al. 2004).
These events have variability amplitudes of 50--400 or more, 
peak \hbox{X-ray} luminosities comparable to those of local Seyfert 
galaxies, soft \hbox{X-ray} spectra, and decay timescales of 
months to years (see Fig.~3a). 
They are observed in both inactive and active galaxies. In inactive 
galaxies the event rate is $\approx 10^{-5}$~yr$^{-1}$, while in 
active galaxies the event rate appears to be $\approx 100$ times higher. 
These outbursts are probably associated with transient fueling events of 
supermassive black holes (SMBHs). Fueling may occur when a star or planet
is tidally disrupted, or in some cases it may be due to accretion-disk
instabilities.

\jdem\ observations should allow the discovery of additional outbursts
from galactic nuclei. The \hbox{X-ray} outbursts above are plausibly expected
to induce accompanying optical variability, and there is direct 
evidence for such optical variability in at least the case of 
IC~3599 (Brandt et~al. 1995; Grupe et~al. 1995). Furthermore,
optical outbursts have been detected from a few galaxies without
near-simultaneous \hbox{X-ray} coverage, including NGC~4552 
(Cappellari et~al. 1999) and perhaps NGC~1068 (de~Vaucouleurs 1991). 
These outbursts should be distinguishable from nuclear supernovae 
based upon their spectral properties and light curves. 

The current data suggest that \jdem\ should detect 2--20 optical 
outbursts from galactic nuclei in its deep, wide-field, and panoramic 
surveys. However, this estimate suffers from significant statistical 
and systematic uncertainties, and \jdem\ observations will allow 
by far the best determination of the frequency of optical outbursts
throughout the Universe. Optical outbursts discovered by \jdem\ 
should be rapidly followed-up with \hbox{X-ray} and multiwavelength 
observatories to determine the nature of these events. Presently, 
the lack of rapid multiwavelength follow-up studies is a major
hindrance to understanding. 


\section{Moderate-Luminosity AGN at High Redshift}

Combined \jdem\ and \hbox{X-ray} studies should allow the discovery of 
hundreds of moderate-luminosity AGN at $z\approx 4$--6;
only $\approx 6$ such AGN are known presently. 
These AGN have luminosities comparable to those of local Seyfert 
galaxies and low-power quasars, and at $z\approx 4$--6 their space 
density is much higher than for the rare, highly 
luminous quasars found by, e.g., the Sloan Digital 
Sky Survey (e.g., Cristiani et~al. 2004). Moderate-luminosity AGN at 
$z\approx 4$--6 can be selected effectively via the application of 
optical/near-infrared ``dropout'' techniques to \hbox{X-ray} sources.
Without complementary \hbox{X-ray} data, many such AGN in the \jdem\ data 
will be difficult and observationally expensive to 
identify as they often have only modest
AGN signatures in the optical/near-infrared. \hbox{X-ray} 
stacking techniques could also be applied to the tens of thousands of 
$z\approx 4$--6 \jdem-discovered galaxies that are not detected 
individually in X-rays. Such stacking would tightly constrain the 
frequency of low-luminosity AGN as well as \hbox{X-ray} emission 
associated with star formation. 

Near-infrared \jdem\ observations of \hbox{X-ray} sources should also break 
the current AGN redshift ``barrier'' at $z\approx 6.5$, allowing the 
discovery of AGN out to $z\sim 10$. A small number of $z>6.5$ 
candidates have already been identified in current deep \hbox{X-ray} surveys
as extreme X-ray-to-optical ratio sources, but most and perhaps all 
probably lie at lower redshifts (Koekemoer et~al. 2004). Wide fields
with sensitive \jdem\ and \hbox{X-ray} coverage would hopefully generate
enough $z\approx 6.5$--10 AGN to elucidate how the black holes formed 
by the deaths of the first stars grew to make luminous quasars at 
$z\approx 4$--6.

\begin{figure}
\includegraphics{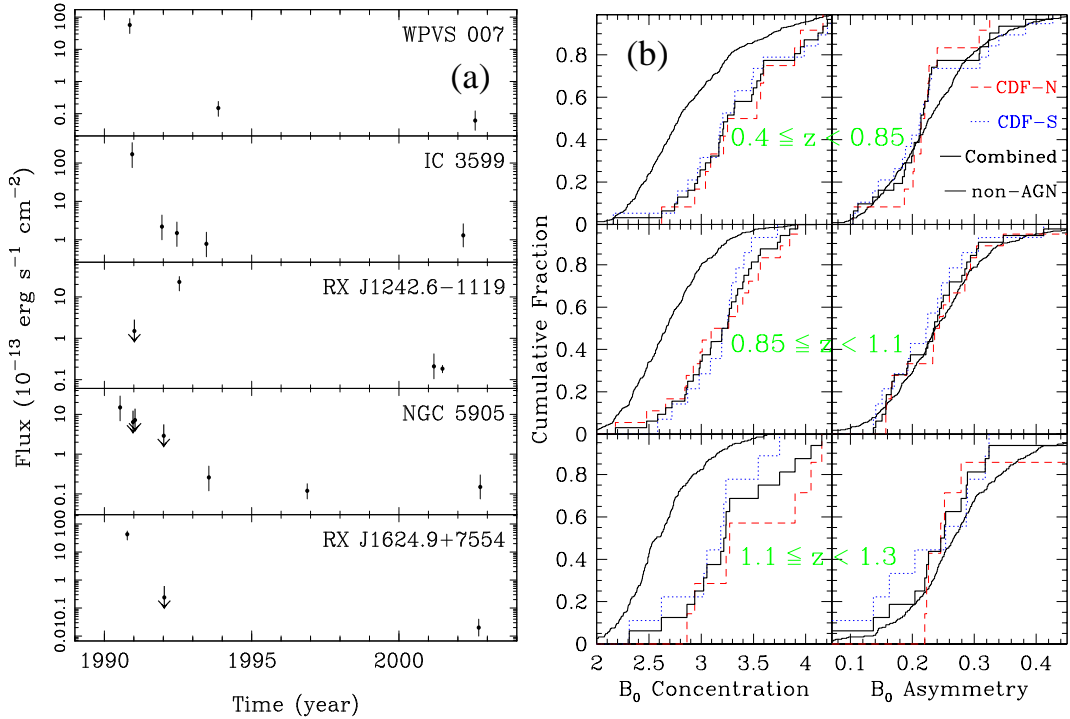}
\vspace{4.0in}
\caption{{\bf (a)} Historical X-ray light curves (from 0.3--2~keV) of 
five large-amplitude X-ray outbursts from galactic nuclei. WPVS~007,
IC~3599, and NGC~5905 appear to be active galaxies, while the other
two appear inactive. From Vaughan et~al. (2004).  
{\bf (b)} Cumulative distribution functions of rest-frame $B$-band
concentration index and asymmetry index for GOODS AGN hosts from
the CDF-N and CDF-S. Cumulative distribution functions for normal
field galaxies are also shown for comparison. Note that AGN hosts
have more concentrated light profiles than normal field galaxies,
but that there is no significant difference in asymmetry. From 
Grogin et~al. (2004).}
\end{figure} 


\section{The Host-Galaxy Morphologies of \hbox{X-ray} Selected AGN}

High-resolution optical and near-infrared imaging, such as that to be 
provided by \jdem, is required to mitigate light blending between many 
moderate-luminosity AGN and their host galaxies, thereby allowing the
hosts to be studied reliably (see \S1). Grogin et~al. (2004) have recently 
studied the host galaxies of samples of 100--200 moderate-luminosity, 
\hbox{X-ray} selected AGN at \hbox{$z=0.4$--1.3} using \hst\ imaging from the Great 
Observatories Origins Deep Survey (GOODS). They find that these AGN are 
preferentially hosted by galaxies with concentrated light profiles,
generally corresponding to more 
bulge-dominated morphologies (see Fig.~3b). Similar
results have been derived from Sersic-model fitting by 
B.D. Simmons et~al., in preparation. Grogin et~al. (2004) suggest
that the locally observed relations between SMBH mass and
host-galaxy properties are already in place at $z\approx 1$
(see, e.g., Graham et al. 2001 for the local relation between 
SMBH mass and concentration). \jdem\ 
imaging would allow these results to be refined using much larger samples,
so that AGN of closely matched luminosity and other properties could
be studied in fine redshift bins. Furthermore, the near-infrared 
capabilities of \jdem\ would allow investigation of SMBH/host-galaxy relations 
at $z\simgt 1.3$, provided sufficient near-infrared sensitivity is achieved 
to overcome surface-brightness dimming. The discovery of an epoch beyond 
which SMBHs and their host galaxies lose their tight relations would help 
to constrain models of SMBH/galaxy co-evolution. 

Grogin et~al. (2004) also investigated if their moderate-luminosity
AGN samples showed enhanced merging or interaction activity relative 
to normal field galaxies, using an asymmetry index and near-neighbor 
counts (see Fig.~3b). They found no 
evidence for enhanced merging or interactions, 
results that are broadly consistent with those found for small samples 
of more luminous quasars (see, e.g., \S7 of Dunlop et~al. 2003). 
Wide fields surveyed by both \jdem\ and \hbox{X-ray} missions would 
provide much larger AGN samples with good source statistics spanning a 
range of at least 10,000 in nuclear luminosity. More subtle differences
in merging and interaction activity might then become statistically
detectable, and differences could be searched for sensitively in 
subsets of the AGN population. 


\section{Some General Prospects for \hbox{X-ray} Support of JDEM}

There are substantial challenges to obtaining wide-field 
\hbox{X-ray} data with sufficient sensitivity to complement \jdem\ properly. 
The exquisitely sensitive imaging of \jdem\ will routinely reach \hbox{$I=27$--30}. 
To obtain reliable \hbox{X-ray} counterparts to optical sources at these faint
flux levels, sub-arcsecond \hbox{X-ray} positions are required. Given the
X-ray-to-optical flux ratios observed for \hbox{X-ray} selected sources
(see Fig.~1), \hbox{X-ray} sensitivities of 
\hbox{$\approx 10^{-17}$--$10^{-16}$~erg~cm$^{-2}$~s$^{-1}$} 
(from \hbox{0.5--2~keV}) are required to complement \jdem\ naturally.\footnote{Note 
that some superb high-energy survey missions currently being
planned, such as \duo\ and \exist, target comparatively bright \hbox{X-ray} 
flux levels that will not naturally match the sensitive optical/near-infrared 
imaging of \jdem. Most of the sources found by these high-energy survey 
missions are expected to have counterparts with $I=18$--23. \jdem\ will 
typically reach flux levels $\sim 1500$ times fainter than these 
counterparts.}
Such sensitivities correspond to \chandra\ exposures 
of 0.2--2~Ms. Thus, the total \chandra\ exposure required
to cover the 15~deg$^2$ \jdem\ deep survey area appropriately would
be \hbox{$\approx 1$--10~yr}; the wide-field ($\approx 500$~deg$^2$) and
panoramic ($\approx 8000$~deg$^2$) \jdem\ surveys would require much
more \chandra\ exposure. Even if \chandra\ survives until the launch
of \jdem\ in $\approx 2014$, \chandra\ time-allocation committees  
may be hesitant to allocate the requisite enormous exposure! 

Future X-ray missions with larger photon collecting areas coupled with
excellent angular resolution, such as ESA's \xeus\ and NASA's \genx, 
offer better prospects for complementing \jdem\ effectively. 
The nominal launch date for \xeus\ of $\approx 2015$ is fairly well
synchronized with that of \jdem, and \xeus\ could obtain appropriate
coverage of the \jdem\ deep survey area in \hbox{$\approx 0.1$--1~yr}. 
\genx, currently under study as a NASA Space Science Vision Mission, 
would probably not begin operation until after $\approx 2020$. It 
could appropriately survey the \jdem\ deep survey area in about a 
week and could also complement the \jdem\ wide-field and panoramic
surveys. 


\section*{Acknowledgments}

I thank 
D.M. Alexander, 
F.E. Bauer, 
N.A. Grogin, 
A.M. Koekemoer,
B.D. Lehmer,  
V.L. Sarajedini, 
B.D. Simmons, 
A.T. Steffen, 
C.M. Urry, and
C. Vignali
for helpful discussions.
Funding from NSF CAREER award AST-9983783 and
CXC grant GO2-3187A is gratefully acknowledged. 



\end{document}